\documentclass[preprint,aps,floatfix,pra]{revtex4-2}

\usepackage[pdftex]{graphicx}
\usepackage{dcolumn}
\usepackage{bm}
\usepackage{color}
\usepackage{xcolor}
\usepackage[shortlabels]{enumitem}

\begin{document}

\title{A fast and strong microactuator powered by internal combustion of hydrogen and oxygen}

\author{Ilia V. Uvarov}
\affiliation{Valiev Institute of Physics and Technology, Russian Academy of Sciences, Yaroslavl Branch, Universitetskaya 21, 150007 Yaroslavl, Russia}
\author{Pavel S. Shlepakov}
\affiliation{Valiev Institute of Physics and Technology, Russian Academy of Sciences, Yaroslavl Branch, Universitetskaya 21, 150007 Yaroslavl, Russia}
\author{Vitaly B. Svetovoy}
\email[Corresponding author: ]{v.svetovoy@phyche.ac.ru}
\affiliation{Frumkin Institute of Physical Chemistry and Electrochemistry, Russian Academy of Sciences, Leninsky prospect 31 bld. 4, 119071 Moscow, Russia}

\begin{abstract}
The development of fast and strong microactuators that can be integrated in microdevices is an essential challenge due to a lack of appropriate driving principles. In this paper, a membrane actuator powered by internal combustion of hydrogen and oxygen in a chamber with a volume of 3.1 nanoliters is demonstrated. The combustion in such a small volume is possible only for an extremely high surface-to-volume (S/V) ratio on the order of 10$^7$~m$^{-1}$. The fuel with this  S/V is prepared electrochemically in a special regime that produces only nanobubbles. A cloud of nanobubbles merges, forming a microbubble, which explodes, increasing the volume 500 times in 10~$\mu$s. The actuator generates an instantaneous force up to 0.5~N and is able to move a body 11,000 times more massive than itself. The natural response time of about 10~ms is defined by the incubation time needed to produce an exploding bubble. The device demonstrates reliable cyclic actuation at a frequency of 1~Hz restricted by the effect of electrolyte aging. After 40,000 explosions, no significant wear in the chamber is observed. Due to a record-breaking acceleration and standard microfabrication techniques, the actuator can be used as a universal engine for various microdevices including micro-electro-mechanical systems, microfluidics, microrobotics, wearable and implantable devices.
\end{abstract}

\maketitle

\section*{Introduction}

Actuators are devices that transform input energy into mechanical movement, such as translation, bending, rotation, or vibration \cite{Li2018}. The development of small, sufficiently powerful, and fast actuators that can be integrated mechanically and electrically in microdevices is one of the key challenges for micro-electro-mechanical systems (MEMS)  \cite{Li2018}, microfluidics \cite{Ma2023}, wearable devices \cite{Leroy2020,Yin2021}, microrobotics \cite{Miskin2020,Yang2020,Bandari2020,Dabbagh2022}, and implantable devices \cite{Yan2022}. Numerous working principles are in use depending on specific applications. Among the most popular transduction mechanisms are electrostatics \cite{Li2002,Tsai2008,Pengwang2016}, piezoelectricity \cite{Ivan2009,Oldham2008,Doll2007,Gao2020,Shan2022}, magnetostatics \cite{Xu2019,Cui2019,Kim2018}, pneumatics and hydraulics \cite{Volder2010,Xavier2022}, electrothermal expansion and electroactive effect \cite{Zeng2015,AbuZaiter2015,Li2015,Tian2021,Chang2019,Tandon2018}, and electroionic actuation \cite{Kim2016,Yu2022,Mahato2023}.

In the ordinary (macroscopic) world, in contrast with microactuators, we use, in most cases, two actuation principles: that of electromagnetic motors and that of internal combustion engines. These two principles allow us to solve most tasks of everyday life. However, both cannot be scaled down to microscopic sizes. For electromagnetic motors, the size is restricted by the dimension of the coil, which is typically larger than a few millimeters  \cite{Chen2015,Qi2023}. Scaling of the internal combustion engines is restricted by the size of the reaction chamber. When the chamber is too small, the heat produced in the reaction escapes too fast through the chamber walls and cannot support the combustion \cite{Lewis1987,Maruta2011,Chou2011}. The smallest volume where normal combustion has been ignited is a two-millimeter bubble containing an acetylene-oxygen mixture \cite{Teslenko2010}. Catalytic combustion in a small volume is possible, but the reaction rate is considerably slower. Realistic actuators using the catalytic process are slow and have dimensions in the millimeter range \cite{Yang2020,Yi2015,Sheybani2015}.

A breakthrough has resulted from a fundamental discovery: the surface of water has been found to strongly accelerate some chemical reactions if the surface-to-volume (S/V) ratio is sufficiently high. The reaction between a ketone and hydrazine to form hydrazone demonstrated strong acceleration in microdroplets compared to the bulk phase \cite{Girod2011}. On the other hand, spontaneous combustion between hydrogen and oxygen in nanobubbles has been discovered \cite{Svetovoy2011}. The observed effects still have no clear explanation, but the subsequent development confirmed the findings \cite{Ruiz2020,Wei2020,Svetovoy2021} and suggested many more accelerated reactions due to "on-water" chemistry.

This finding removed the fundamental restriction on the minimum volume in which combustion can occur and opened a principal way for the fabrication of internal combustion engines with all three dimensions in a microscopic range \cite{Svetovoy2014}. The actuator that used the combustion of hydrogen and oxygen in nanobubbles has been demonstrated \cite{Uvarov2018}. It has a chamber 8~$\mu$m high and 500~$\mu$m in diameter covered by a 30-$\mu$m-thick polydimethylsiloxane (PDMS) membrane with electrodes inside and filled with an electrolyte. Short ($\sim 1\; \mu$s) voltage pulses of alternating polarity applied to the electrodes electrochemically generate H$_2$ and O$_2$ nanobubbles (NBs) in the chamber that push the membrane up. When the pulses are switched off, the NBs with different gases disappear rapidly in the surface-assisted combustion reaction, and the membrane returns to its initial position. This actuator provides a stroke of the membrane of about 10~$\mu$m at the amplitude of the applied voltage, of about 10~V, and has a response time in the range of 10~ms. However, it cannot be considered a true internal combustion engine since it uses combustion to quickly remove the gases, rather than rapidly expand the volume.

Nanobubbles are difficult for direct observation as separate entities \cite{Alheshibri2016,Ma2022}, but optical effects induced by many NBs show that the electrodes are covered by a cloud of NBs \cite{Postnikov2018}. If the amplitude of the voltage pulses exceeds a threshold value (typically $11-13$~V), the concentration of NBs in the cloud is so high that they touch each other and merge \cite{Svetovoy2021,Svetovoy2020}. A microbubble (MB) emerges filled with a stoichiometric mixture of H$_2$ and O$_2$ gases and nanodroplets from the space between the NBs. These nanodroplets provide a high S/V ratio necessary for the spontaneous combustion of the gases. The combustion occurs in a few tens of nanoseconds \cite{Svetovoy2020}; the MB grows about 10 times in size, then shrinks and disappears. The entire process takes only $200 - 300\ \mu$s, indicating that the explosion of an MB is observed. Exploding MBs are better candidates for microscopic internal combustion engines. In this case, the fuel is prepared electrochemically in the form of H$_2$ and O$_2$ NBs; the ignition occurs when the concentration of NBs reaches a critical value; the mixture explodes and expands providing mechanical work; the water vapor condenses; and the system returns to the initial state without any exhaust.

In this paper, a microactuator that uses internal combustion as a driving principle is presented. After an incubation time of 10~ms, an MB explodes in the chamber and expands the volume 500 times in only 10~$\mu$s. Fast inflation generates a large acceleration and, therefore, a large instantaneous force, which is able to move bodies with a mass thousands of times larger than that of the working chamber. The fabrication method is compatible with standard microtechnologies. For this reason, the actuator can be built into various microsystems, such as MEMS, microfluidic systems, microrobots, and micromanipulators. The actuator chamber is 16~$\mu$m high and has a diameter of 500~$\mu$m, but the fabrication process has good potential for further reduction of all dimensions. A mixture of hydrogen and oxygen NBs with a high S/V ratio is used as a fuel. This mixture is generated electrochemically, but different methods for preparing fuel may be explored in the future. Like its macroscopic counterpart, the actuator can be used to provide various mechanical movements. 

\section*{Results and Discussion}

\begin{figure}[ptb]
\begin{center}
\includegraphics[width=1.0\textwidth]{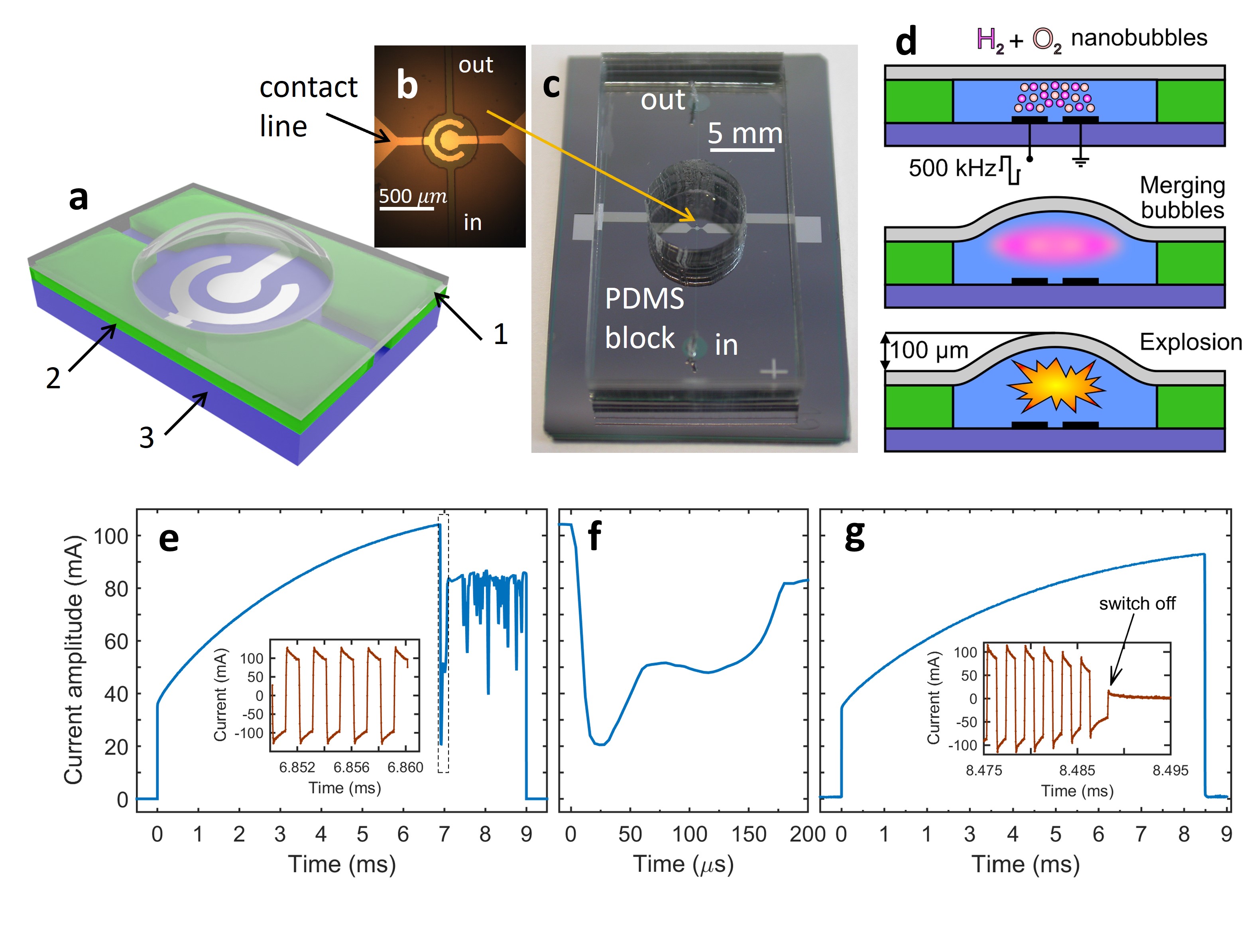}
\caption{\textbf{Actuator.} \textbf{a} Schematic view of the actuator where 1 is the PDMS layer (30~$\mu$m thick), 2 is the SU-8 layer (16~$\mu$m thick), and 3 is the oxidized Si wafer. \textbf{b} Photo of the working chamber with concentric electrodes. The electric signal is applied to the contact lines and the chamber is filled with the electrolyte through the filling channels marked as in and out. The chamber is sealed by a transparent PDMS membrane (nearly invisible).  \textbf{c} The macroscopic framing of the actuator used for wiring and tubing. The contact pads and in/outlets are visible. The well that opens access to the chamber is 8~mm in diameter. \textbf{d} Schematic presentation of the processes taking place in the chamber during actuation. \textbf{e}  Amplitude of the current averaged over 4 pulses (two periods) as a function of time. The inset shows several current pulses prior to the explosion. \textbf{f} Current behavior near the major explosion indicated by the dashed rectangular in  e. \textbf{g}  Current in the chamber when using the automatic shut-off system. The inset shows the last pulses. \label{fig:fig1}}
\end{center}
\end{figure}

The actuator is shown schematically in Fig.~\ref{fig:fig1}a. The chamber ($16\ \mu$m high and $500\ \mu$m in diameter) is made of an SU-8 negative photoresist covered by a 30-$\mu$m-thick PDMS membrane. Concentric metallic electrodes are located at the bottom of the chamber and made of Ru, which demonstrated long-time stability in the alternating polarity electrolysis \cite{Uvarov2022}. The chamber is filled with a molar solution of Na$_2$SO$_4$ through the filling channels. The top view of the chamber is presented in Fig.~\ref{fig:fig1}b where one can also see in/outlet channels and contact lines. The entire device shown in Fig.~\ref{fig:fig1}c is macroscopic for convenience of handling, but the actuator is a small chamber in its center with a volume of 3.1~nL. The structure of the device is similar to that described earlier \cite{Uvarov2022}. The processes occurring in the chamber during actuation are illustrated schematically in Fig.~\ref{fig:fig1}d. Application of alternating polarity pulses with a frequency of 500~kHz generates a cloud of NBs covering the electrodes. At a certain moment, the NBs in the most densely packed area of the cloud coalesce. The resulting MB then explodes, propelling the membrane upwards with great force.


The idea of using the surface-assisted explosions of MBs for actuation was expressed in \cite{Uvarov2018}, but the scale of possible improvement was not recognized. Moreover, a controlled actuation with exploding microbubbles was deemed impossible because of parasitic explosions. The problem is explained in Fig.~\ref{fig:fig1}e where the amplitude of the current through the electrolyte is shown. The process is driven by rectangular voltage pulses of alternating polarity with an amplitude of 12~V and frequency 500~kHz. The current changes with the same frequency as shown in the inset on a timescale of $10\ \mu$s. When the pulses are switched on, the amplitude of the current starts to increase. The increase is due to heat production in occasional recombination of H$_2$ and O$_2$ NBs rather than the Joule heating \cite{Svetovoy2014}. The growth continues up to 6.9~ms (incubation time) and then the current suddenly drops down in only $10\ \mu$s. The current stays low for about $200\ \mu$s, as one can see in Fig.~\ref{fig:fig1}f, and then fluctuates strongly. A major fall in the current is explained by the explosion of an MB formed between the electrodes. This phenomenon has been described for open systems (electrodes covered by the electrolyte, no chamber) where expanding MBs have been directly observed with fast cameras \cite{Postnikov2016,Svetovoy2020}. However, in open systems, no current fluctuations after the major explosion have been observed. The current fluctuations are interpreted as uncontrolled explosions of smaller MBs that appear in the confined space already saturated with the NBs. In an unconfined electrolyte layer, the liquid saturated with the NBs leaves with the first explosion and the process starts from the beginning in fresh liquid. The fluctuations do not allow the device to be used for controlled actuation.

\subsection*{Timing of a stroke}

The control electronics are adjusted to exclude parasitic explosions. For this purpose, the current amplitude averaged over a few periods is monitored, and the pulses are switched off when the amplitude decreases to a certain level as shown in Fig.~\ref{fig:fig1}g. This simple modification provides the actuator with very useful properties. First, only one explosion at a time drives the actuator in a controlled way, reproducing the mechanism of internal combustion. Second, in comparison with normal actuation regime, the stroke of the membrane is 10 times larger and reaches $100\ \mu$m. The volume of the chamber increases more than three times, up to 9.7~nL. Third, and most importantly: after an incubation period on the order of 10~ms, the total lifting time of the membrane up to 100~$\mu$m is only 20~$\mu$s.

In the open system, it has been demonstrated that the sharp current decrease is accompanied by the growth of an exploding MB that covers the electrodes and blocks the current \cite{Svetovoy2020}. In a confined space, the process is developed similarly but the timing and some other aspects are different. The timing of the explosion in the chamber is well described by the curve presented in Fig.~\ref{fig:fig1}f. The steep section describes the expansion of MB after the explosion. The current drops from a constant value to its minimum in 20~$\mu$s, but the linear decrease lasts about 10~$\mu$s and is considered as a characteristic time. It corresponds to the growth of the bubble, which fills the entire chamber and pushes the membrane up. At a minimum, the current is not exactly zero since a certain amount of liquid remaining in the corner between the bottom and the wall of the chamber allows the current to flow. When the current is minimal, the bubble has a maximum size, and the membrane has a maximum deflection. The bubble then shrinks for 35~$\mu$s to a minimal value (maximum on the curve) where its size is not as small as the size of the initial MB. This stage and the subsequent disappearance of the bubble are explained by cavitation, which is less pronounced than in the open system \cite{Svetovoy2020}. This fact is important for the wear of the actuator and will be discussed below.


\begin{figure}[ptb]
\begin{center}
\includegraphics[width=1.0\textwidth]{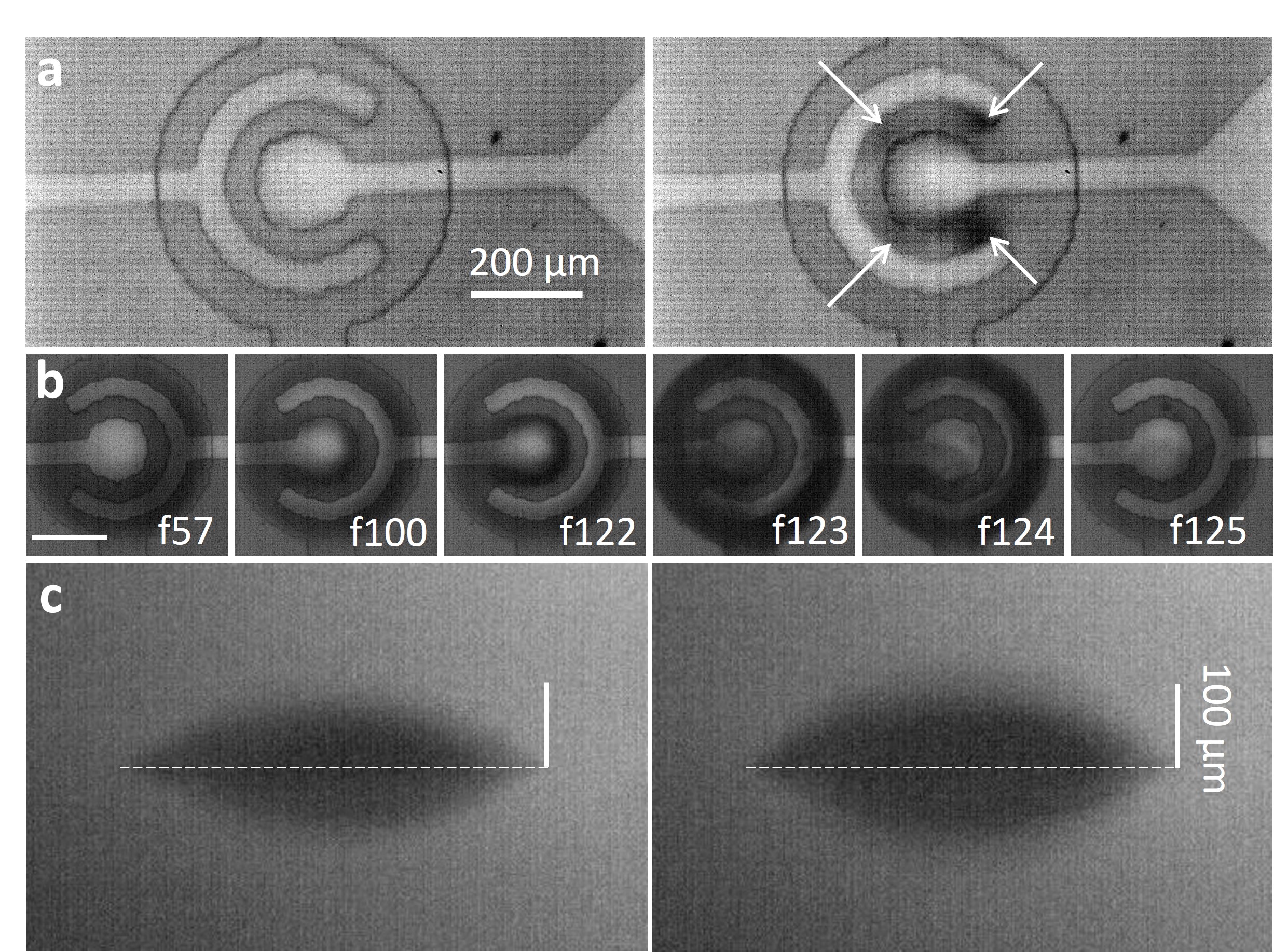}
\caption{\textbf{Processes in the chamber.} \textbf{a} (Left) Image of the chamber before application of the pulses to the electrodes. (Right) The same chamber 8~ms after switching on the pulses with the amplitude 11.5~V, slightly below the threshold. The arrows indicate the positions of dense spots in the cloud of NBs. \textbf{b} Series of frames from Supplemental Video~S1 made at the amplitude of pulses 12.5~V, slightly above the threshold. The explosion in the chamber is between frames 122 and 123. The white bar in the first image is $200\ \mu$m. \textbf{c} Side view of the membrane deflection induced by the explosion in the chamber. The angle between the light direction and the substrate is $8^{\circ}$. The thin dashed line is the position of an undeflected membrane. The two images correspond to separate explosions. \label{fig:fig2}}
\end{center}
\end{figure}


The MB filled with a stoichiometric mixture of gases and nanodroplets is the fuel that drives the actuator. The initial size of this bubble defines the energy released in the combustion process. Since the timescales for the merging of NBs and the following combustion are $\sim 10$~ns \cite{Svetovoy2020}, it is difficult to observe the initial bubble directly. One can only indirectly estimate its size. Figure~\ref{fig:fig2}a shows the top view of the chamber before application of the voltage pulses (left image) and the same view in 8~ms after application of the pulses of alternating polarity (right image) slightly below the explosion threshold. The darker spots indicated by the arrows show dense areas in the clouds of NBs, but the darker rim on the central electrode results from membrane deflection due to the presence of many NBs in the chamber.
With the increase of the density of NBs in the cloud (at higher voltage amplitude), the NBs start to merge, forming an MB. From the right image, one can estimate the largest dark spot as having a diameter of 40~$\mu$m and the height is assumed to be 16~$\mu$m (height of the chamber), hence the volume of the spot is $V_b=20$~pL. In contrast to conventional combustion, the reaction within this MB is initiated spontaneously and proceeds without significant temperature rise \cite{Svetovoy2021}.

Figure ~\ref{fig:fig2}b shows the dynamics of the chamber as a series of images made by a fast camera (see Supplemental Video~S1) at 10,000 frames-per-second (fps) for a shutter time of $100\ \mu$s. The first image f57 shows the initial state of the chamber. In frame f100, which is taken at time $t=4.35$~ms, one can see that the membrane is slightly deflected and a dense area of NBs is formed between the central electrode and the lower end of the encircling electrode. The membrane deflection and the dense area are well pronounced in frame f122. Between this and the next frame (f123), the initial bubble explodes. The process is too fast to see the details, but it is clear that the membrane deflects significantly. In the next frame (f124), the current is already zero, but the liquid is moving violently in the chamber. In frame f125, the membrane practically returns to its initial state and fluid dynamics in the chamber slows down. Supplemental Fig.~S1 shows the amplitude of the current as a function of time and demonstrates the position of each frame from Fig.~\ref{fig:fig2}b in this graph.

Significant rise of the membrane occurs at a timescale of $10\ \mu$s. It is too fast to observe the membrane deflection with an interferometer. The stroke of the membrane has been estimated by taking pictures from the side synchronized with the explosion. The minimal possible delay between the current stop and the shot was $70\ \mu$s, and for this reason the shutter time was increased to $50\ \mu$s to collect light from the highest position of the membrane. Although this blurs the position, it is possible to evaluate the rise of the membrane close to its maximum. Figure~\ref{fig:fig2}c shows the images of two separate explosions. In the left image the stroke is about $100\ \mu$m and in the right image it is $120\ \mu$m.

\subsection*{Energetics of a stroke}


\begin{figure}[ptb]
\begin{center}
\includegraphics[width=1.0\textwidth]{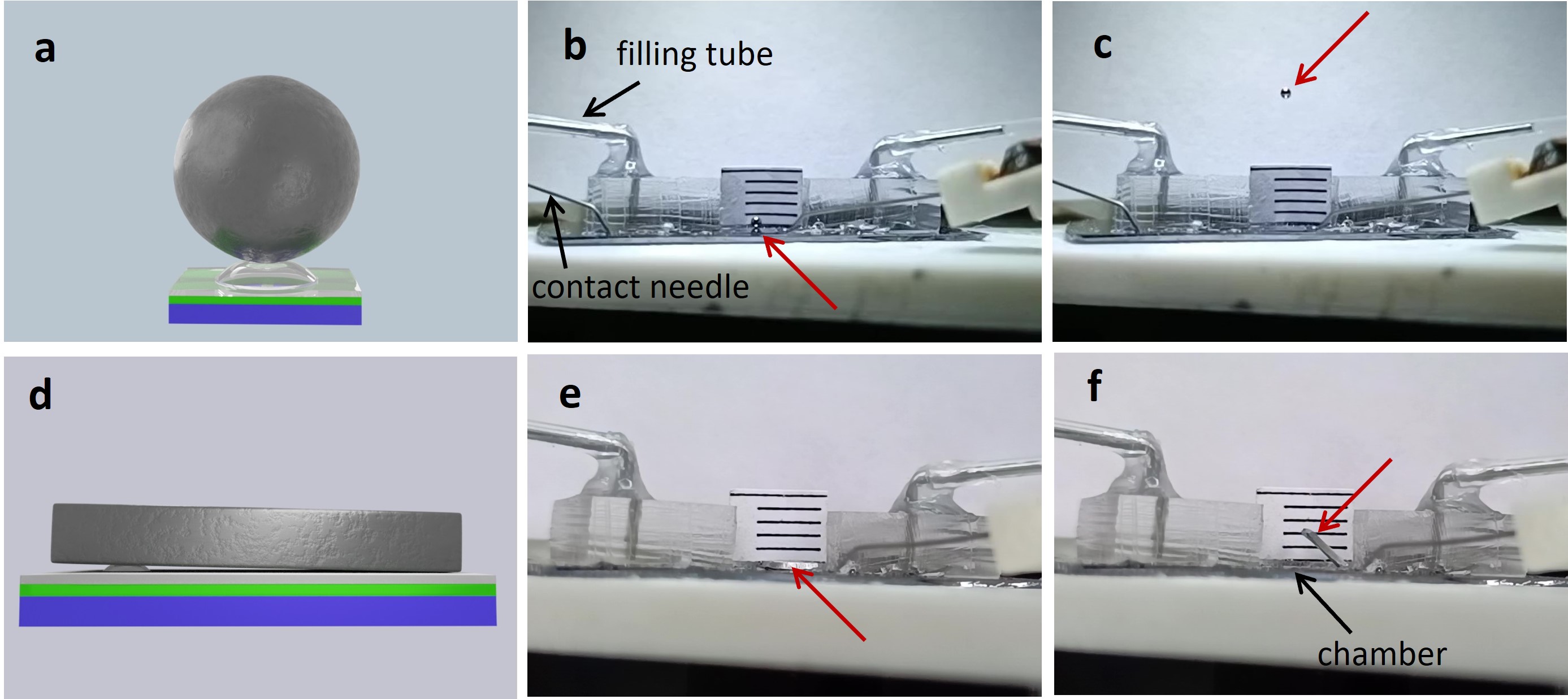}
\caption{\textbf{Strikes on macroscopic bodies.} \textbf{a} Schematic view of the ball on top of the chamber (relative dimensions are preserved). \textbf{b} A metallic ball with a mass of 4.1~mg on top of the chamber. A thick PDMS rectangular bar (see Fig.~\ref{fig:fig1}c) is cut along the longer side to open the chamber for side views. The distance between the black horizontal lines drawn on the paper is 1~mm. The red arrow shows the position of the ball at rest. \textbf{c} The ball, indicated by the red arrow, was pushed by the actuator to the highest point of 12~mm. \textbf{d} Schematic view of the plate on top of the chamber (relative dimensions are preserved). \textbf{e} The left end of a metallic plate with a mass of 35~mg is placed on top of the chamber, and the right end lies on the substrate. The plate is shown by the red arrow. \textbf{f} An explosion in the chamber raises the left end of the plate to 2~mm. The position of the plate is shown by the red arrow and that of the chamber is shown by the black arrow. \label{fig:fig3}}
\end{center}
\end{figure}


It is worth stressing once more that a stroke of $100\ \mu$m is reached in $20\ \mu$s so that the average velocity of the rising membrane is about 5~m/s. The maximum velocity is even larger and estimated as 10~m/s since the main growth of the bubble occurs for $10\ \mu$s. The ratio of the latter two values allows one to estimate the initial acceleration developed by the membrane, which is as large as $10^6$~m/s$^2$. When the actuator is loaded, the acceleration and, therefore, the instantaneous force will be smaller but still very significant. To make sure of this, simple but convincing experiments have been performed. A metallic sphere with a diameter of 1.05~mm and a mass of $m=4.1$~mg was released from a ballpen and mounted on top of the chamber. This is schematically shown in Fig.~\ref{fig:fig3}a, where a realistic relationship between the sizes of the ball and the chamber is used. Figure~\ref{fig:fig3}b shows the ball in the initial position. Square voltage pulses with an amplitude of 12.3~V and a frequency of 500~kHz applied to the electrodes led to an explosion in the chamber, which hit the ball. Figure~\ref{fig:fig3}c shows the ball at the highest point at a height $H=12$~mm. In this case, the actuator transfers the mechanical energy $E=mgH = 0.48\ \mu$J to the ball, where $g=9.8$~m/s$^2$ is the free-fall acceleration. The same energy is equal to the kinetic energy of the ball at the initial point of the trajectory. In this way, one can calculate the initial velocity of the ball as $v=(2gH)^{1/2} = 0.49$~m/s. Its acceleration is evaluated as $a \approx v/t_0=4.9\times 10^4$~m/s$^2$, where $t_0=10\ \mu$s is the characteristic time, and the instantaneous force is $F=ma\approx 0.20$~N. Supplemental Text~S1 provides some details of the calculations. Thus, a microactuator with a volume of 3.1~nL is able to produce significant velocity and apply a very large force to a macroscopic ball that is nearly 200 times larger in volume and 1,300 times more massive.

It is difficult to place the ball exactly in the center of the membrane, keeping in mind the adhesion forces between the ball and membrane, but the precision of the positioning defines the efficiency of the energy transfer.  A more successful strike is shown in Video~\ref{vid:video_1} where the ball leaves the frame range. The video was made with a smartphone at a frame rate of 240 fps. The amplitude of the pulses was 12.5~V, which is slightly higher than in the previous case. Neglecting the resistance of air, one can evaluate from the video the highest point of the trajectory as $H=29$~mm. This value corresponds to the energy transfer $E=1.2\ \mu$J with initial velocity $v=0.76$~m/s and  instantaneous force $F\approx 0.31$~N. This energy can be considered as a lower limit of the work done by the actuator: $W>1.2\ \mu$J, because only a small area of the membrane is in direct contact with the ball. The diameter of the contact area is estimated as $41\ \mu$m assuming  Hertzian contact \cite{Johnson1985} and using Young's modulus $E_Y=1.4$~MPa and Poisson ratio $\nu=0.48$ as PDMS parameters \cite{Liu2009}. The average current amplitude corresponding to Video~\ref{vid:video_1} was shown in Fig.~\ref{fig:fig1}g, where the inset shows the current near the time of explosion. The incubation period before the explosion is 8.5~ms.


\begin{video}
\includegraphics[width=0.5\textwidth]{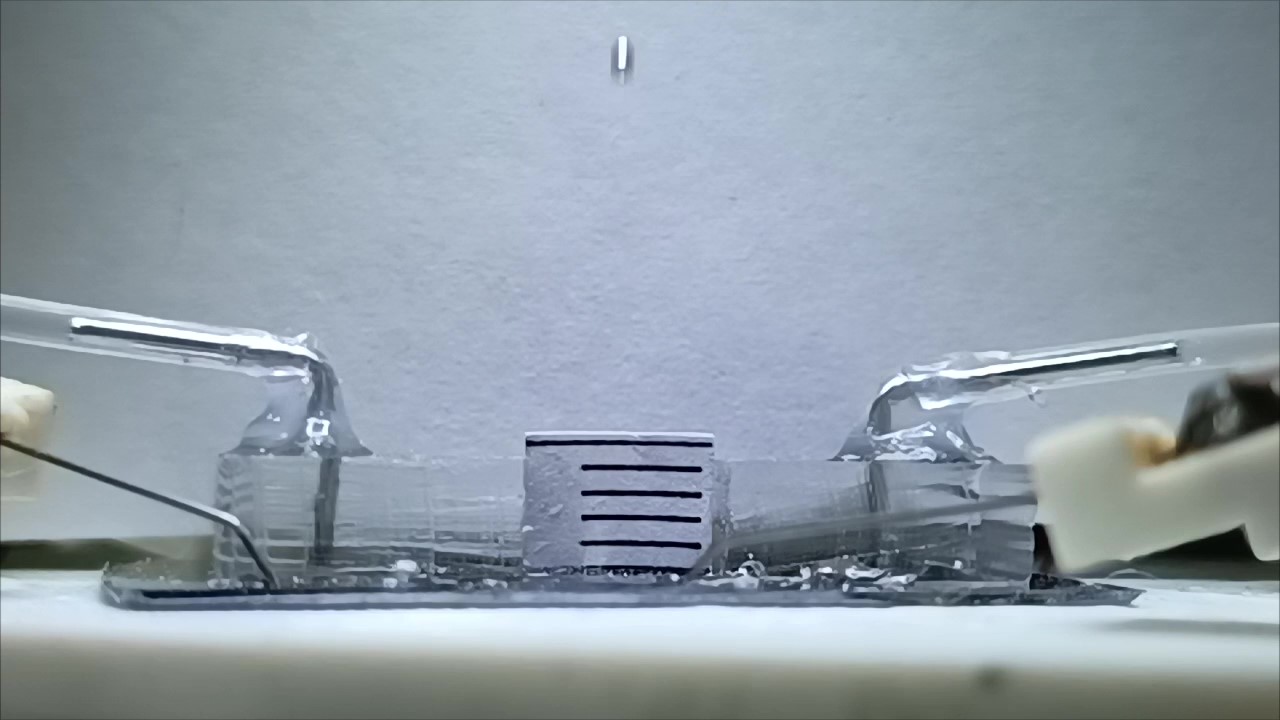}
\caption{\textbf{Strike of the ball video.} A frame rate of 240 fps is insufficient for sharp images. The highest point $H=29$~mm is reached outside the frame.\label{vid:video_1}}
\end{video}


On the other hand, one can find an upper limit on the work produced by the actuator. The initial MB filled with a stoichiometric mixture of gases and nanodroplets drives the actuator. The volume of this bubble was estimated as $V_b=20$~pL. This MB emerges when NBs are  packed so densely that they nearly touch each other. In such a bubble, a part of the volume occupied by gas is $f_0=0.74$ (densely packed spheres) and the rest, $1-f_0=0.26$, is filled by nanodroplets providing a high S/V ratio in the MB. If it is assumed that all NBs have the same radius $r$, then, before the combustion reaction, the pressure in the bubble contains a large Laplace contribution:  $P_b=P_a+2\gamma/r$, where $\gamma$ is the surface tension of liquid and $P_a$ is the ambient pressure. The energy, produced by combustion of gases in this MB, is
\begin{equation}\label{eq:comb_energy}
  E_r = \frac{2P_bV_b}{3\mathcal{R}T}\left|\Delta H_r\right|f_0,
\end{equation}
where the number of molecules in the bubble is estimated using the gas law and $\mathcal{R}$ is the gas constant, $T$ is the room temperature, and $H_r=-242$~kJ/mol is the enthalpy of the combustion reaction. Taking the bubble radius in the range $r=30-40$~nm \cite{Postnikov2018}, one finds that the reaction energy is $E_r=3.5-4.7\ \mu$J. However, not all  this energy is able to produce a  mechanical work. Part of it dissipates in the surrounding liquid and solid materials.

To see how the actuator behaves at a significantly higher load, a metallic plate with a mass of $M=35$~mg  and dimensions of $3.9\times 2\times 0.6$~mm has been used. Note that the mass of the plate is 11,000 times larger than that of the chamber. The plate was mounted with the left end on the actuator, but the right end rested on the substrate as shown schematically in Fig.~\ref{fig:fig3}d (relative dimensions are preserved). The initial situation is demonstrated in Fig.~\ref{fig:fig3}e. The pulses with a smaller amplitude (12~V) were applied to the electrodes, which increased the incubation time to 13~ms. After the explosion, the left end rose by 2~mm as one can see in Fig.~\ref{fig:fig3}f. A video of the process is presented in Supplemental Video~S2. In this case, one has to follow the plate's center of mass, which is raised by the explosion to a height of $H=1$~mm. The energy received by the plate is $E=MgH =  0.34\ \mu$J and the initial velocity of the center of mass is $v=(2E/M)^{1/2} = 0.14$~m/s. The exploding bubble grows to its maximum size with the same characteristic time, $t_0=10\ \mu$s, but the maximum bubble size is smaller for the loaded system. It can be seen in Supplemental Fig.~S2, where the current near the point of explosion is shown for loaded and unloaded cases. Thus, for the instantaneous force one finds $F=Mv/t_0\approx 0.49$~N. The work done by the actuator to move the plate is smaller than for the ball. The difference can be related to a smaller size of the exploding bubble at its maximum for the actuator with a heavy load.

\begin{figure}[ptb]
\begin{center}
\includegraphics[width=0.5\textwidth]{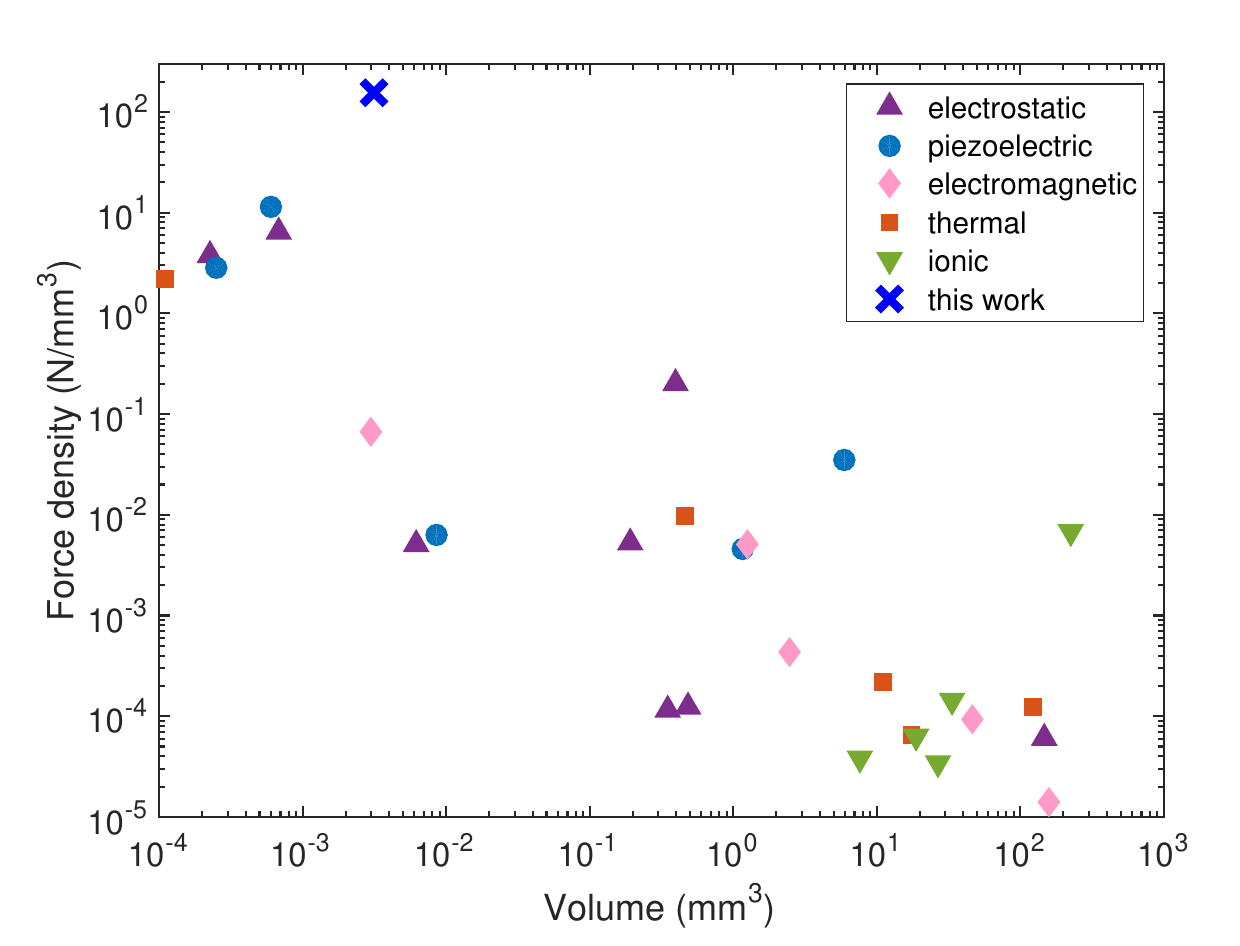}
\caption{\textbf{Comparison of different types of actuators.} The force density versus effective volume. The data for this comparison are taken from various sources, including \cite{Li2002,Patel2012,Khan2014,Felder2015,Ba_Tis2015,Contreral2017,Schaler2018,Abbasalipour2022} for electrostatic, \cite{Oldham2008,Conway2007,Ivanov2012,Xie2015,Jafferis2015} for piezoelectric, \cite{Wright1997,Feldmann2007,Choi2012,Wang2021,Han2021} for electromagnetic, \cite{Zeng2015,Li2015,Sinclair2000,Yosefi2014,Garces2019} for thermal, and \cite{Kim2016,Yu2022,Mahato2023,Wu2015,Chen2017} for ionic actuators.  \label{fig:comparison}}
\end{center}
\end{figure}

A high force generated by the actuator makes it stand out from a row of similar devices. Figure~\ref{fig:comparison} shows the force density produced by the actuators that use different physical principles as a function of the effective volume of the actuator. One can see that the exploding bubble actuator generates the highest force density and is one of the smallest devices. Furthermore, because it is possible to reduce the volume of the chamber further, this actuator could become the smallest and strongest.


Electrostatic, piezoelectric, and thermal actuators can generate the highest force densities of several Newtons per cubic millimeter, but they all have low displacement capabilities. The electrostatic scratch drive actuator moves in steps of no more than 25~nm \cite{Li2002}, while the MEMS switch crab-leg electrode deflects by 550~nm \cite{Patel2012}. Piezoelectric drives have a somewhat higher stroke of several micrometers \cite{Oldham2008,Ivanov2012}. The thermal chevron-type actuator \cite{Sinclair2000} achieved a maximum displacement of 14~$\mu$m, which is also a low value compared to the stroke of the exploding bubble actuator. Additionally, the power consumption for this thermal actuator is very high. Thus, the proposed actuator develops more than one order of magnitude higher force density combined with a significantly larger displacement.

\subsection*{Cyclic operation}

\begin{figure}[ptb]
\begin{center}
\includegraphics[width=1.0\textwidth]{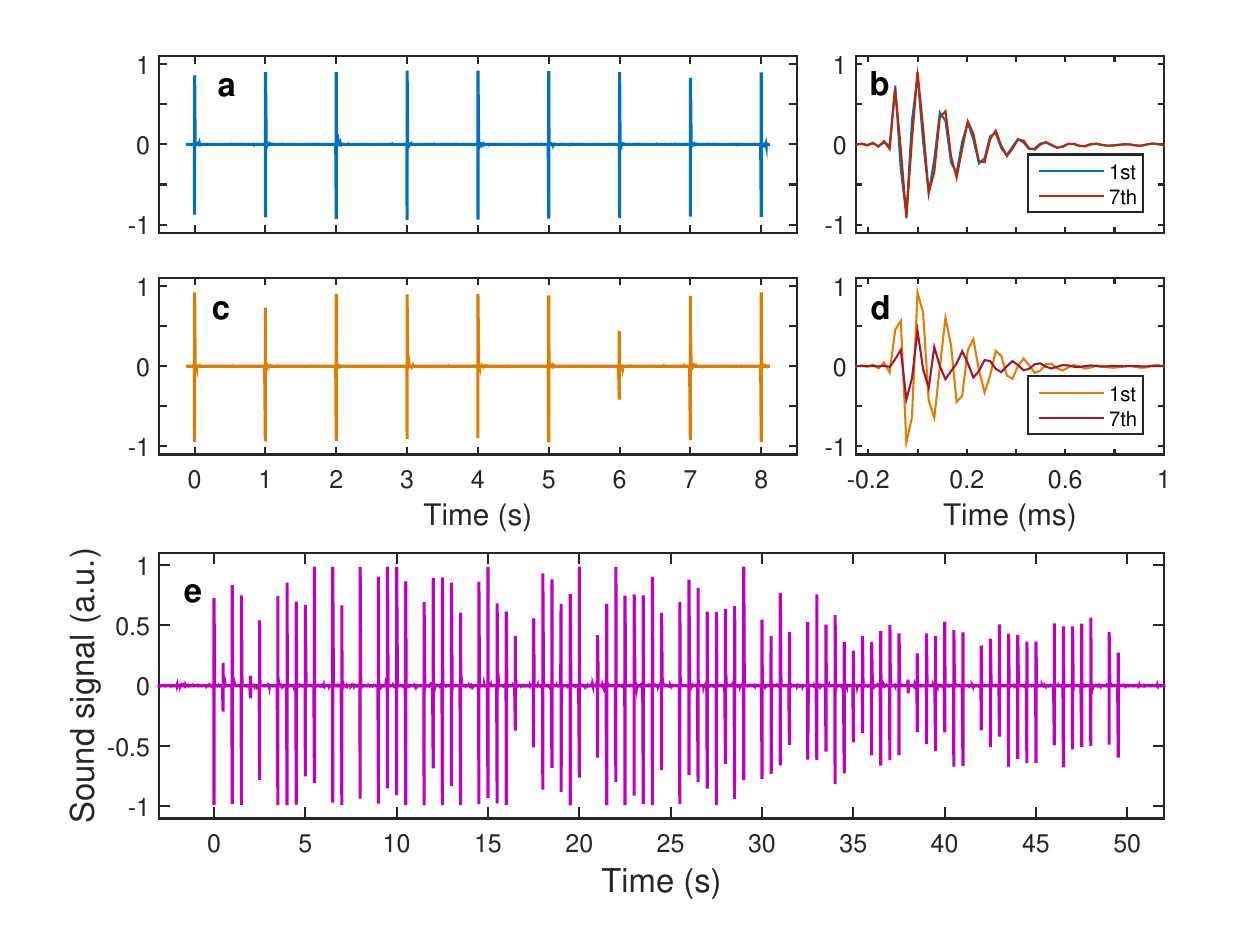}
\caption{\textbf{Actuation in cyclic regime.} \textbf{a} Audio signal of a series of nine explosions at a cyclic frequency of 1~Hz.  \textbf{b} Zoomed view on the first and seventh lines in panel a. Time is counted from the maximum of the line. \textbf{c} An additional series of explosions at a frequency of 1~Hz. The seventh event has a smaller amplitude.  \textbf{d} A zoomed view of the first and seventh lines in panel c. \textbf{e} Audio signal of a series of 100 explosions at a frequency of 2~Hz. The figure demonstrates some irregularities. \label{fig:fig4}}
\end{center}
\end{figure}

There are many applications where actuators have to be used in a cyclic regime, as, for example, in micropumps or in engines for flying or crawling microrobots.  Therefore, the actuator has been tested for periodic explosions in the chamber. The high acceleration produced by the actuator is a source of sound that is audible to the ear. This sound is a convenient way to monitor the explosions in the chamber. The sound of nine explosions at a cyclic frequency of 1~Hz is presented in Supplemental Audio~S1. The number of explosions is restricted by the equipment's ability to write the current simultaneously with the sound. The corresponding sound signal is shown in Fig.~\ref{fig:fig4}a. It consists of nine narrow lines with a similar width and amplitude. Each line has a width of about 0.5~ms, and zoomed views for two of them are presented in Fig.~\ref{fig:fig4}b. As one can see, all explosions look similar. The sound is presented by the main frequency at 11~kHz. Of course, the acceleration is the source of a wide range of frequencies, but the membrane plays the role of a filter. The first eigenfrequency of the membrane is evaluated as 9.8~kHz with all the higher modes being outside the audible range \cite{Azimi1988}. This value is found for a $30\ \mu$m-thick layer of PDMS with the parameters $E_Y=1.4$~MPa and $\nu=0.48$  \cite{Liu2009}. This frequency is somewhat smaller than the observed value, but the stressed membrane has a larger Young modulus as measured in \cite{Liu2009}; it will increase the lowest eigenfrequency.


However, from time to time, explosions of smaller intensity appear, as one can see in Fig.~\ref{fig:fig4}c where the 7th line has a smaller amplitude than the others. The detailed structure of this line is slightly different from the regular one as shown in Fig.~\ref{fig:fig4}d, but the difference can be explained by digitizing errors. A distinctive feature of the events with a smaller amplitude is a shorter incubation time. If the regular explosions in Fig.~\ref{fig:fig4}c  are characterized by incubation times of $12.2\pm 0.4$~ms, the 7th line has an incubation time of 8.1~ms. It is natural to assume that a smaller amplitude corresponds to a smaller initial MB that explodes. Such an event is expected if in liquid there is a preferable place for formation of a dense spot in the cloud of NBs. The nature of these defects needs a separate investigation, but one could expect that O$_2$ and N$_2$ gases dissolved initially in the electrolyte can play this role. After a number of explosions in the chamber, these gases could appear in the form of relatively stable MBs playing the role of seeds for dense spots of H$_2$ and O$_2$ NBs.

In the case of normal actuation (no explosions), the NBs produced for 20~ms disappear completely from the chamber in 80~ms \cite{Uvarov2018}. Actuation with a cyclic frequency of 10~Hz in the conditions involving MB explosion produce explosions with uncontrolled amplitude. Figure \ref{fig:fig4}e demonstrates the problem with a long actuation series (100 explosions) with a frequency of 2~Hz.
It appears that some explosions missed, but it is because their amplitude is too low ($\sim 10^{-3}$) to be resolved on the vertical scale of the figure. The other specificity is a slow reduction of the average amplitude with time. These two features have different origins. The first one shows that the electronics do respond even on rather weak explosions. This can be improved by using a more sophisticated algorithm to switch off the current. The second problem involves the aging of the electrolyte and needs more detailed investigation. It is possibly related to the dissolved gases in the electrolyte before the experiments.

\subsection*{Wear of the actuator}

The problem of electrolyte aging will not be discussed futher, but it is clear that it can be resolved, at least technically, by delivering fresh solution automatically from a built-in reservoir. However, it is known that in normal actuation regime, the electrodes are subject to severe physical and chemical wear \cite{Svetovoy2011,Svetovoy2021,Uvarov2022}. Relatively soft materials such as Au, Cu, or Pt are physically dispersed by highly energetic reaction between NBs near the electrodes. More stiff materials, such as Ti or W, are oxidized or dissolved electrochemically, reducing the current supported by the electrodes in a few minutes. For normal actuation regime, Ru electrodes demonstrated unique properties among many other materials \cite{Uvarov2022}. Ruthenium electrodes are oxidized, but ruthenium oxide is a stiff conductive material in contrast with titanium oxide, which is a poor conductor. In the exploding actuation regime, Ru electrodes have not yet been tested.

\begin{figure}[ptb]
\begin{center}
\includegraphics[width=1.0\textwidth]{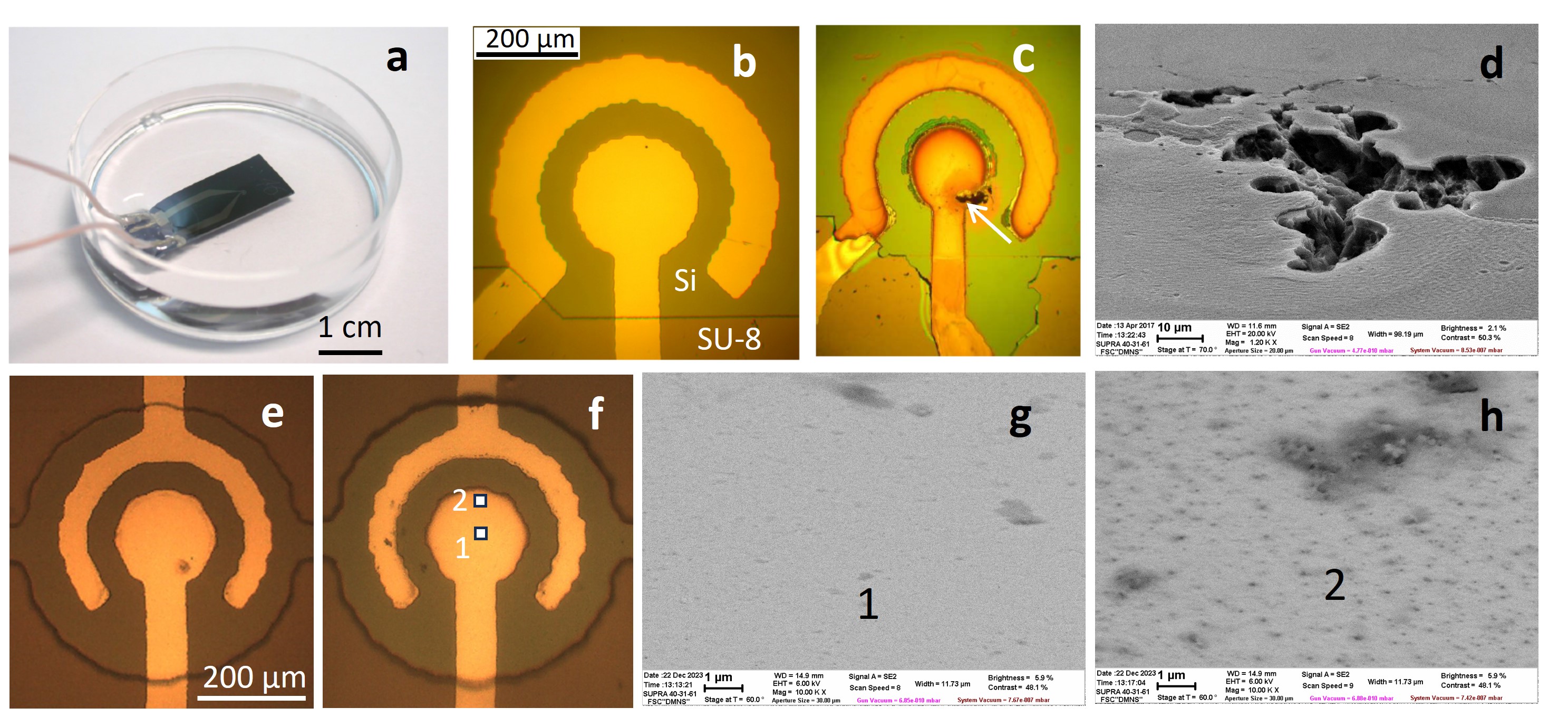}
\caption{\textbf{Wear of the actuator.} Damage in open system (top row). \textbf{a} A sample with electrodes in a Petri dish filled with the electrolyte.  \textbf{b} Electrodes before the process. \textbf{c} Electrodes after 4,000 explosions. The arrow indicates the area of damaged substrate. \textbf{d} SEM image of the arrowed area in c. Damage in the chamber (bottom row). \textbf{e} Chamber with electrodes before the process.  \textbf{f} The chamber after 40,000 explosions. White squares show the areas investigated with SEM. \textbf{g} SEM image of the area 1 taken at a stage angle of 60$^{\circ}$. \textbf{h} SEM image of area 2. \label{fig:fig5}}
\end{center}
\end{figure}

In open systems, exploding MBs can be very destructive. In this case the growth of the exploding MB is not confined. It can grow to a size as large as 1.2~mm \cite{Svetovoy2020} and then shrink to a very small size in the same place where the initial MB exploded. This is the cavitation process that generates shock waves damaging the underlying surface \cite{Brennen1995}. Figure~\ref{fig:fig5}a-d shows how serious the cavitation damage can be in an open system. Special samples containing the electrodes on a Si wafer have been prepared. The contact lines are covered by SU-8 for insulation. The sample in a Petri dish filled with the electrolyte is shown in Fig.~\ref{fig:fig5}a. Panel b shows the electrodes before the process. Application of alternating polarity pulses with an amplitude slightly above the threshold generates explosions. The electrodes after 4,000 explosions are shown in Fig.~\ref{fig:fig5}c. One can see considerable damage to the insulating layer and some destruction of the electrodes. Moreover, a black spot indicated by the arrow shows that even a hard material like Si is also damaged. This is clear from the SEM image in Fig.~\ref{fig:fig5}d.


A completely different situation is observed in a confined space of the actuator, as Fig.~\ref{fig:fig5}e-h demonstrates. Panel e shows the electrodes before the process. The electrodes after 40,000 explosions in the chamber are shown in Fig.~\ref{fig:fig5}f. Only small changes in the color of electrode edges are visible. No damage to the substrate, chamber walls, or covering membrane can be seen. Inspection with SEM of the spots marked as 1 and 2 shows that the central part of the central electrode (Fig.~\ref{fig:fig5}g) was not affected by the explosions at all. The edge of the central electrode is slightly changed (Fig.~\ref{fig:fig5}h), but this is due to the reaction between NBs during the incubation period. There is some difference in chemical composition in areas 1 and 2: more oxygen is observed in area 2 (see Supplemental Table~S1 and explanations there).

The significant difference between open and confined systems is explained by the influence of two factors. First, in the confined case, the initial microbubble is restricted in growth. For this reason, the energy of the collapsing bubble is smaller than in the open system. Second, the initial bubble explodes not in the center of the chamber but between the electrodes. Because of this asymmetry, the collapsing bubble is not focused, and its energy is distributed over a wider area of space. Thus, the actuator with ruthenium electrodes has high durability not only in normal actuation regime \cite{Uvarov2022} but also in the regime with exploding MBs.

\subsection*{Power consumption}

The main power consumed by the actuator is spent on fuel preparation.
To generate an initial microbubble, one has to decompose water. The amount of gas must be very large because of high pressure inside NBs. Only 18\% of the produced NBs are distributed in the volume of the chamber; the rest disappear in the combustion reaction in a thin layer above the electrodes \cite{Uvarov2022}. These are the main reasons for low energy conversion efficiency. However, only a little additional energy is needed to turn the volume of the chamber saturated with NBs into the oversaturated state where the exploding microbubbles are formed spontaneously. As an example, one can consider the event shown in Video~\ref{vid:video_1}. The electrical energy needed to produce the explosion as 7.3~mJ.
On the other hand, the useful mechanical work delivered by the actuator to the ball is 1.2~$\mu$J. Thus, the total energy conversion efficiency is $1.6\times 10^{-4}$.

However, it is not assumed that the electrical energy is supplied to the actuator continuously. If, for example, the same event is produced once per second, then the consumed power is 7.3~mW. This is a reasonable value for microactuators. The consumed power will decrease with the decrease of the chamber size because one needs less energy to saturate a smaller volume with NBs. For example, the very first actuator operated in normal regime had a chamber with dimensions $100\times 100\times 5$~$\mu$m  \cite{Svetovoy2014}. A stroke close to the explosion threshold (see Fig. S4 in \cite{Svetovoy2014}) consumes 23~$\mu$J of electrical energy, which is 317 times less than for the present actuator.

The power consumption problem may be significantly reduced if a way can be found to prepare the fuel by a direct mixture of gases and then to provide the conditions for a high S/V ratio to ignite the mixture. This could involve injecting nanodroplets in the chamber with a gas mixture, or passing a mixture of gases through a wetted porous membrane, or something else. In any case, surface-assisted combustion of gases in a microchamber demonstrates outstanding results for fast and strong actuators, as this work shows.

\section*{Methods} 

\subsection*{Fabrication}

A standard 100~mm Si wafer is thermally oxidized in wet oxygen; the SiO$_2$ layer has a thickness of 1~$\mu$m. The metallic electrodes and contact lines are formed on this layer by magnetron sputtering. A 10-nm-thick adhesive Ti layer is followed by a 500-nm-thick Al layer needed to reduce the ohmic losses; finally 150~nm of Ru is deposited. The side walls of the working chamber and filling channels are made of the photoresist SU-8 3005. The post-exposure bake is performed at a relatively low temperature of 95$^\circ$~C to keep residual epoxy groups on the SU-8 surface \cite{Zhan2011}.

A liquid compound Sylgard 184, with a 10:1 mixing ratio is applied to a separate wafer by spin-coating and curing at 100$^\circ$~C for 35 min to form a 30-$\mu$m-thick PDMS layer. A separate PDMS block 4~mm thick with an 8-mm-diameter through hole is bonded to the layer on the Si wafer by plasma treatment of both surfaces. The block is then separated from the plate, while the through hole remains closed by the membrane. The inlet and outlet holes with a diameter of 0.5~mm are punched.

The final step is sealing the working chamber with the membrane by bonding the PDMS block to the SU-8 surface. For this purpose, the PDMS is exposed for 1~s to nitrogen plasma at a pressure of 0.5~mbar and a power of 700~W. The exposed surface is brought in contact with the SU-8 surface, and the assembly is heated to 100$^\circ$~C. As a result, the amino groups generated on the PDMS surface using plasma interact with the residual epoxy groups on the SU-8 surface. All the fabrication steps are shown in Supplemental Fig.~S3.

\subsection*{Characterization}

The setup to investigate the performance of the actuator is schematized in Supplemental Fig.~S4. The actuator is mounted on a 3D-printed platform \cite{Uvarov2023} and filled with the electrolyte using a syringe and a microfluidic tube of 0.8~mm inner diameter (see Supplemental Fig.~S5). The driving signal is applied to the actuator via tungsten probes installed on the contact pads. The signal is provided by a homemade pulse generator that automatically switches the pulses off upon  explosion. It is connected to a computer with the governing software. The voltage and current flowing through the electrodes are recorded by a USB oscilloscope PicoScope 5442B. The actuator is observed from the top using an optical microscope equipped with a video camera Moticam 1SP. Fast processes in the chamber are recorded by a high-speed camera Photron FASTCAM 1024PCI. The videos of the ball and plate moved by the actuator are made using a smartphone TECNO Camon 20 Pro 5G at 240~fps. The sound produced by the actuator is recorded by a microphone Huawei AM115 in a separate notebook to exclude crosstalk. The distance between the microphone and the membrane is 4~mm.

To measure the membrane deflection, the platform is rotated such that the membrane is observed from the side. The membrane deflection is registered by a camera Jenoptic ProgRes CF triggered by the pulse generator. A sync pulse is applied to the camera when the driving signal of the actuator is switched off upon the explosion. The delay between the sync pulse and the start of shooting is 70~$\mu$s due to camera performance, while the exposure time can be adjusted. The minimum shutter time is 20~$\mu$s.

Special samples for experiments in an open system are fabricated in the same way as the electrodes for the actuator. Damage to the electrodes is investigated optically and with a scanning electron microscope (SEM) Zeiss Supra 40. Chemical composition of the electrodes is determined using an energy dispersive X-ray spectrometer Oxford Instruments INCA x-ACT installed in the SEM. Chemical analysis is performed at an accelerating voltage of 6 kV.

\section*{Conclusions}

The actuator presented in this paper is impossible from the point of view of normal combustion. In a small volume of the working chamber, the combustion cannot be ignited and supported since the heat escapes too rapidly via the walls. Surface-assisted combustion is a fundamentally new regime involving a high S/V ratio on the order of 10$^7$~m$^{-1}$. It can be called "cold combustion" because it is ignited spontaneously at room temperature and characterized by a low temperature rise. The detailed mechanism of this phenomenon and that of "on-water" chemistry in general is still not clear, but the gas-water interface seems to generate free radicals in some way \cite{Svetovoy2021}.

Explosion of an MB in the chamber ensures a very high acceleration of the membrane, which is a characteristic feature of this device. None of the existing actuators can produce a comparable acceleration. High acceleration means a large force developed by the actuator. The microscopic actuator demonstrated a force as large as 0.49~N moving a macroscopic plate 11,000 times more massive than the working chamber. These characteristics make it possible to use the actuator as an engine in various microsystems.

The total response time of the actuator by its nature is on the order of 10~ms. The cyclic actuation at the moment is restricted by a frequency of 1~Hz. This is related to the aging of the electrolyte -- a phenomenon that needs more investigation to eliminate it, but a pure technical solution to the problem is also possible. Ruthenium electrodes do not show any significant degradation in the long-term operation when the actuator works in normal regime (without explosions) \cite{Uvarov2022}. Explosions could produce cavitation damage as happens in open systems. However, it was demonstrated here that the cavitation is not destructive because of the asymmetry of explosions in a confined space, and no significant wear of the actuator is observed after 40,000 explosions.

The fuel preparation step is the main source of power consumption. One has to produce a high density of NBs in the chamber electrochemically and only a small part of them is used to produce mechanical work. Still, for cyclic operation, power consumption stays in a reasonable range below 10~mW. In order to reduce power consumption and make the actuator an autonomous device, it would be interesting to explore a different approach to preparing the fuel with a high S/V ratio.

\begin{acknowledgments}
I.V.U. and P.S.S. acknowledge support from Program No. FFNN-2022-0017  and V.B.S. from Program FFZS-2022-0015 of the Ministry of Science and Higher Education of Russian Federation.
\end{acknowledgments} 


%

\end{document}